\documentclass[10pt,preprint]{aastex}

\newcommand{\Sc}{{\cal S}}
\newcommand{\Ec }{{\cal E}}

\shorttitle{Cosmic Rays from Gamma Ray Bursts in the Galaxy}
\shortauthors{Dermer \& Holmes}

\begin{document}

\title{Cosmic Rays from Gamma Ray Bursts in the Galaxy}

\author{Charles D. Dermer\altaffilmark{1}, Jeremy M.\ Holmes\altaffilmark{2,3}}
\altaffiltext{1}{E\,O.\,Hulburt Center for Space Research, Code 7653,
Naval Research Laboratory, Washington, DC 20375-5352; e-mail: 
dermer@gamma.nrl.navy.mil}
\altaffiltext{2}{Thomas Jefferson High School for Science and Technology, 
6560 Braddock Road, Alexandria, VA 22312, USA}
\altaffiltext{3}{Present address: Florida Institute of Technology, 150 W. University Blvd., Melbourne, FL 32901 }

\begin{abstract}
The rate of terrestrial irradiation events by galactic gamma-ray
bursts (GRBs) is estimated using recent standard-energy results.  We
assume that GRBs accelerate high-energy cosmic rays, and present results
of three-dimensional
simulations of cosmic rays moving in the Galactic magnetic field and
diffusing through pitch-angle scattering. An on-axis GRB extinction
event begins with a powerful prompt $\gamma$-ray and neutron pulse,
followed by a longer-lived phase from cosmic-ray protons and
neutron-decay protons that diffuse towards Earth. Our results force a
reinterpretation of reported $\sim 10^{18}$ eV cosmic-ray anisotropies
and offer a rigorous test of the model where high-energy cosmic rays
originate from GRBs, which will soon be tested with the Auger Observatory.
\end{abstract}


\section{Introduction }

Observations link GRBs---brief flashes of $\gamma$-ray light emitted
by sources at cosmological distances---with star-forming galaxies and
high mass stars \citep{pwk00}. These results indicate that GRBs are
produced by a rare type of supernova where the evolved core of a
massive star collapses to a black hole. Normal supernovae, where the
stellar core collapses to form a neutron star, are thought to produce
nonrelativistic supernova remnant shocks that accelerate cosmic rays
with energies $\lesssim 10^{14}$ eV.  The origin of higher energy
cosmic rays is controversial. One possibility is that high-energy
($\gtrsim 10^{14}$ eV) cosmic rays are accelerated by the 
relativistic shocks formed by GRB explosions in the Galaxy and 
throughout the universe \citep{vie95,wax95,der02,wda04}, 
dating back to earlier suggestions of a Galactic origin of 
high-energy cosmic rays \citep{kfk69}.

If GRBs accelerate high-energy cosmic rays, then Galactic GRBs could
be detected as cosmic-ray sources from neutron $\beta$-decay emissions
\citep{ikm04}. GRBs with their jets oriented toward the Earth
have potentially lethal consequences, and may have contributed to past
extinction episodes \citep{dls98,mel04}.  The sources of 
the $\approx 10^{18}$ eV
cosmic-ray excesses reported from measurements made with the SUGAR (Sydney
University Giant Air Shower Recorder) and AGASA (Akeno Giant Air
Shower Array) detectors (\citet{bel01,hay99}; see \citet{nw00} for review)
have been proposed to result from past GRBs in
the Galaxy \citep{bie04}.

To investigate these ideas, the rate of GRB events at different fluence
levels is estimated, based on recent findings about beaming in GRBs
\citep{fra01,bfk03}. A 3D propagation model is used to simulate the 
sequence of irradiation events that occurs when a GRB jet is pointed
towards Earth. The results suggest that a GRB jet could have produced
radiations that contributed to the Ordovician extinction event
\citep{mel04}. Our results are contrary to the claim that the SUGAR excess
could be produced by a GRB in the Galaxy. If the Auger Observatory
confirms the
SUGAR cosmic-ray point source, then a model of high-energy cosmic rays 
from galactic GRBs is incomplete.

\section{Rate of GRB Events at Different Fluence Levels}

Most of the electromagnetic radiation from a GRB is emitted during the
prompt and early afterglow phases on timescales of minutes to
hours. We define the bolometric photon fluence $\varphi
=\Sc\varphi_{\odot}$ with reference to the Solar energy fluence
$\varphi_{\odot} = 1.4\times 10^{6}\Sc$ ergs cm$^{-2}$ received at
Earth in one second.  Significant effects on atmospheric chemistry
through formation of nitrous oxide compounds and depletion of the
ozone layer is found when $\Sc \gtrsim 10^2$ -- $10^3$
\citep{rud74,tho95,geh03,tho05}, taking into account the very hard incident
radiation spectrum of GRBs. Reprocessing of incident GRB radiation
into biologically effective 200 -- 320 nm UV radiation (i.e., with
$1/e$ lethality) on eukaryotes occurs when $\Sc \gtrsim 10$ -- 10$^2$
\citep{sw02,ssw04}.

Achromatic beaming breaks in GRB optical/IR afterglow light curves
\citep{sta99}, if due to jetted GRBs, imply typical GRB jet opening
half-angles $\langle \theta_j \rangle \cong 0.1$. Analyses
\citep{fra01,bfk03} show that long-duration GRBs have a standard total
energy $\Ec \cong \theta_j^2 E_{\gamma,iso}/2
\cong 10^{51}\Ec_{51}$ ergs, with $\Ec_{51} 
\cong 1.33$, a variance by a factor of $2.2$, and a low $E_{\gamma,iso}$
population.  Here $E_{\gamma,iso}$ is the apparent isotropic
$\gamma$-ray energy release inferred directly from observations. If
the cone of emission from a GRB at distance $R$ intercepts the
line-of-sight to Earth, then the radiant fluence is given by $\varphi
= E_{\gamma,iso}/4\pi R^2$. Thus the maximum sampling distance $R_s$
of a GRB with apparent isotropic $\gamma$-ray energy release
$E_{\gamma,iso}$ to be detected at the fluence level $\varphi >
\varphi_{th}=\Sc\varphi_{\odot}$ is
\begin{equation}
R_{s} = \sqrt{{E_{\gamma,iso}\over 4\pi \varphi_{th}}}\cong
{1.1 {\rm~kpc}\over (\theta_j/0.1)}\sqrt{{\Ec_{51}\over (\Sc/10^3)}}\;.
\label{rsamp}
\end{equation}

When $\Sc \lesssim 10^5$, the sampling distance to a typical GRB
exceeds the $\approx 100$ pc disk scale height of molecular clouds and
OB associations, and we can approximate the distribution of GRBs in
the Galaxy by a uniform disk of radius $R_{MW} \cong 15R_{15}$
kpc. The fluence size distribution of GRBs in this approximation is
given by
\begin{equation}
\dot N(>\Sc ) \simeq \dot N_{GRB} 
\;({R_s\over R_{MW}})^2\; {\cal P}(\theta_j)\;,
\label{nsc}
\end{equation}
where $\dot N_{GRB}$ is the rate of GRBs in the Milky Way, and ${\cal
P}(\theta_j)\cong \theta_j^2/2$ is the probability that the Earth lies
within the emission cone of a 2-sided jet.  If GRBs follow the star
formation rate history of the universe, then one GRB occurs every
$10^5 t_5$ years in the Milky Way
\citep{wda04,der02}, with $t_{5} \simeq 0.1$ -- 1.  Thus
\begin{equation}
\dot N(>\Sc ) \simeq {0.3\over R_{15}^2} \;
{\Ec_{51}\over (\Sc/10^3) t_5}\; {\rm Gyr}^{-1}\;.
\label{nsc_1}
\end{equation}
Estimates (\ref{rsamp}) and (\ref{nsc_1}) show that a GRB at a distance
$\approx 1$ kpc with $\Sc \gg 10^2$ takes place about once every
Gyr, and more frequently if $t_5\cong 0.1$.
  
\section{Cosmic Ray Propagation in the Galaxy}

We have developed a numerical model where cosmic rays move in response
to a large-scale magnetic field that traces the spiral arm structure
of the Galaxy, and diffuse through pitch-angle scattering due to
magnetic turbulence. The magnetic field $\vec B$ of the Galaxy is
modeled as a bisymmertric spiral for the Galaxy's disk, and a dipole
magnetic field for the Galaxy's halo \citep{aes02}. The evolution of
the particle momentum $\vec p = m\gamma\vec \beta c$ is found by
solving the Lorentz force equation $d\vec p/dt = q \vec
\beta\times \vec B$, where $q$ and $m$ are the particle's charge and
 mass, respectively, $\vec \beta c$ is its velocity, and $\gamma =
(1-\beta^2)^{-1/2}$.

Magnetic turbulence causes a particle to change its pitch angle by
$\approx \pi/2$ when travelling the mean-free-path $\lambda$. The
energy dependence of $\lambda$ is obtained by extrapolating the 
expression for $\lambda$
in a diffusion-model fit \citep{wda04} to the measured ionic flux near the
knee of the cosmic-ray spectrum \citep{kam01} to high energies. Our approach 
assumes
isotropic turbulence that is uniform in the disk and halo 
of the Galaxy; anisotropic turbulence is more realistic \citep{gs97}, but
increases the number of free parameters in the model. The
particle's azimuth and cosine angle are randomly chosen between $0$
and $2\pi$ and between $\mu_{min}$ and 1, respectively, after every
$\lambda [(1-\mu_{min})/2]^2/c\Delta t$ steps, where $\Delta t$ is the
time interval of each step in the numerical integration, which is set
equal to a small fraction of the local gyroperiod.

The propagation calculations are performed in the test-particle
approximation, assuming that the cosmic rays do not  affect their surrounding environment.
The energy density of the cosmic-ray shell can, however, greatly exceed the magnetic field
energy density of the ISM, in which case the protons will sweep up material, causing 
strong adiabatic losses and rapid deceleration withn a Sedov length scale of $\sim$ pc. 
This forms a shock and a second phase of cosmic ray acceleration, which will require
a hydrodynamic simulation to treat. The cosmic-ray neutrons decay over the 
entire jet volume. Comparing the ISM magnetic-field energy density
with the energy density of the neutron-decay protons over
their decay volume  shows that the propagation 
of neutrons with 
\begin{equation}
\gamma \gtrsim {2\times 10^{7}\over (B/3\mu{\rm G})^{0.62} (1-\cos\theta_j)^{0.31}}\;.
\label{gamma_neutron}
\end{equation} 
Thus our subsequent discussion
of  $\sim 10^{18}$ eV cosmic-ray neutrons is accurately described in the test particle approximation.

The propagation of cosmic-ray protons and neutrons, with and without
the effects of diffusive scattering, is illustrated in Fig.\ 1.
Cosmic-ray neutrons travel $\approx 10 (E/ 10^{18}$ eV) kpc before
decaying. Cosmic-ray protons with energies $\gtrsim 3\times 10^{18}$
eV and cosmic-ray neutrons with energies $\gtrsim 10^{18}$ eV escape
almost directly into intergalactic space. Protons with energies
$\lesssim 3\times 10^{17}$ eV diffusively escape from the Galaxy
through a combination of Larmor motions and pitch-angle scatterings.

Fig.\ 2 displays the cosmic-ray halo that surrounds a GRB source $\sim
4\times 10^{11}$ s after the event. The GRB is modeled by
radially-oriented jets with $\theta_j = 0.1$. A conical shell with
effective angular extent larger than 0.1 radians forms as a result of
directly accelerated protons and neutron-decay protons with $E \gtrsim
10^{18}$ eV. The turbulent wave spectrum that is resonant with these
high energy cosmic rays is poorly known. In a simulation with
pitch-angle scattering absent, energy-dependent features and wall-like
structures are formed, as shown in the inset to Fig.\ 2.

\begin{figure}
\includegraphics[scale=1.0]{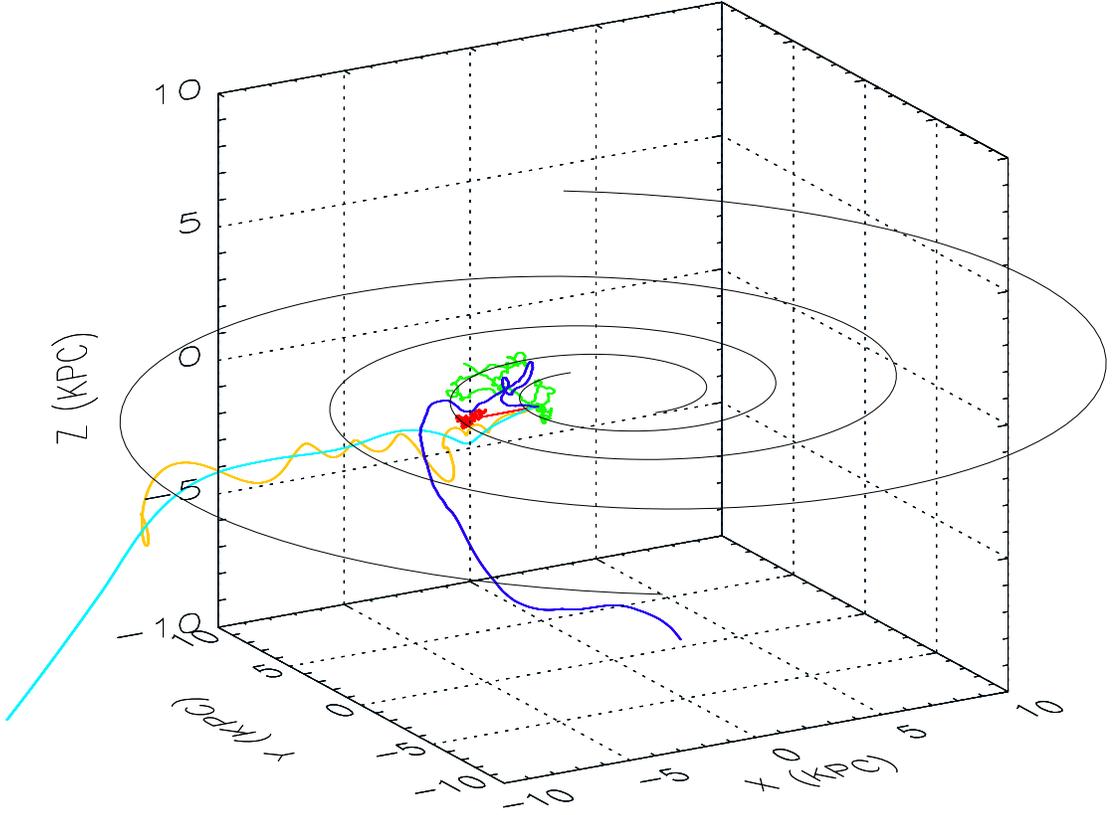}
\caption{Trajectories of cosmic rays ejected outward from a 
cosmic-ray source located at 3 kpc from the center of the Galaxy.  The
spiral curves show the peak magnetic field values in the disk of the
galaxy for the magnetic field model of \citet{aes02}. The orange
and turquoise curves show trajectories taken by cosmic-ray protons
with Lorentz factors $\gamma = 10^9$ and $3\times 10^9$, respectively,
with diffusive scattering omitted, and the green and blue curves
illustrate the effects on protons with respective energies when
diffusive pitch-angle scattering is included. The red curve shows the
path taken by a cosmic-ray neutron with $\gamma = 3\times 10^8$ and
mean decay length of 3 kpc, including scattering effects on the path
of the neutron-decay proton. }
\label{fig1}
\end{figure}
\clearpage

\begin{figure}
\includegraphics[scale=0.5]{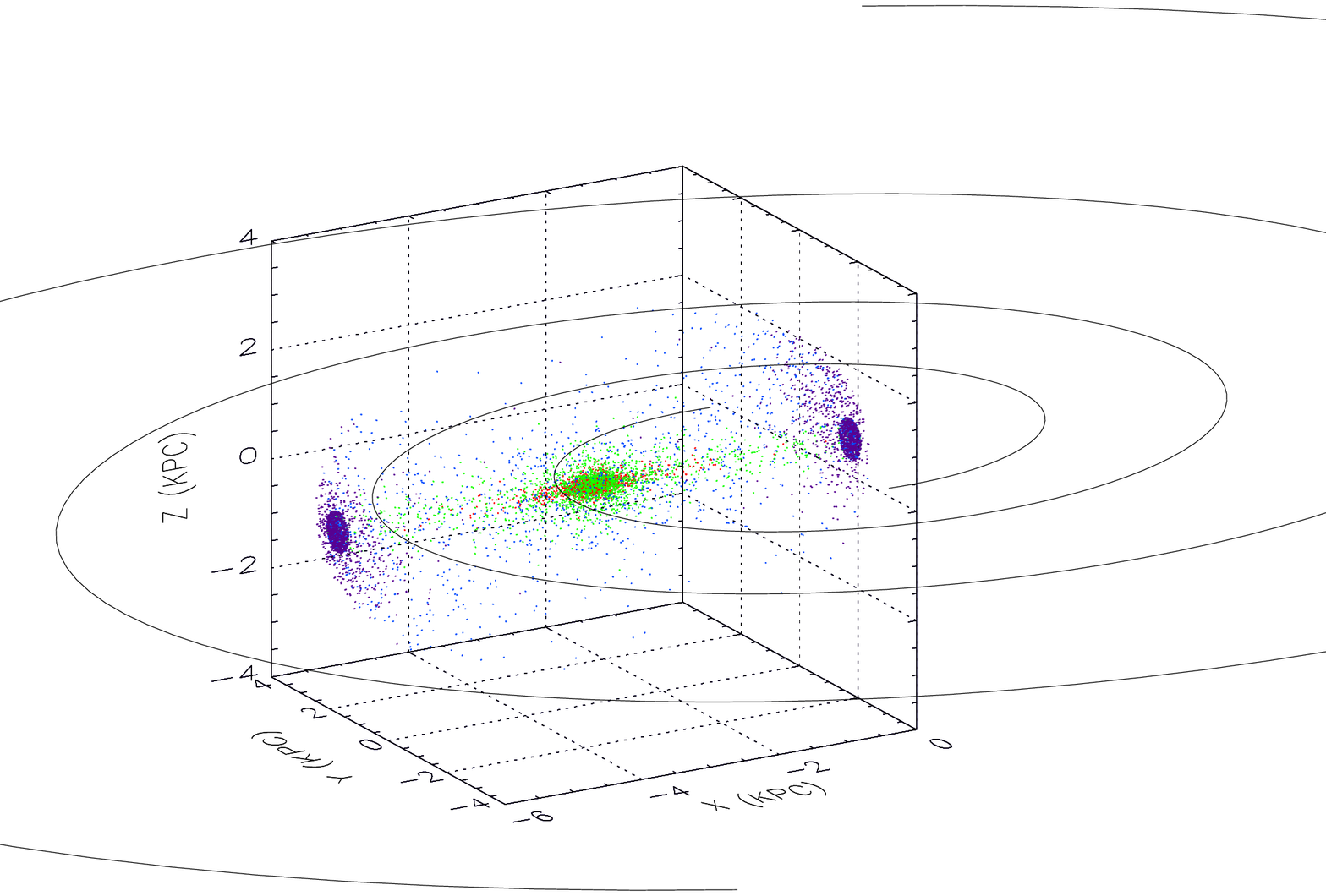}
\vskip0.2in
\includegraphics[scale=0.5]{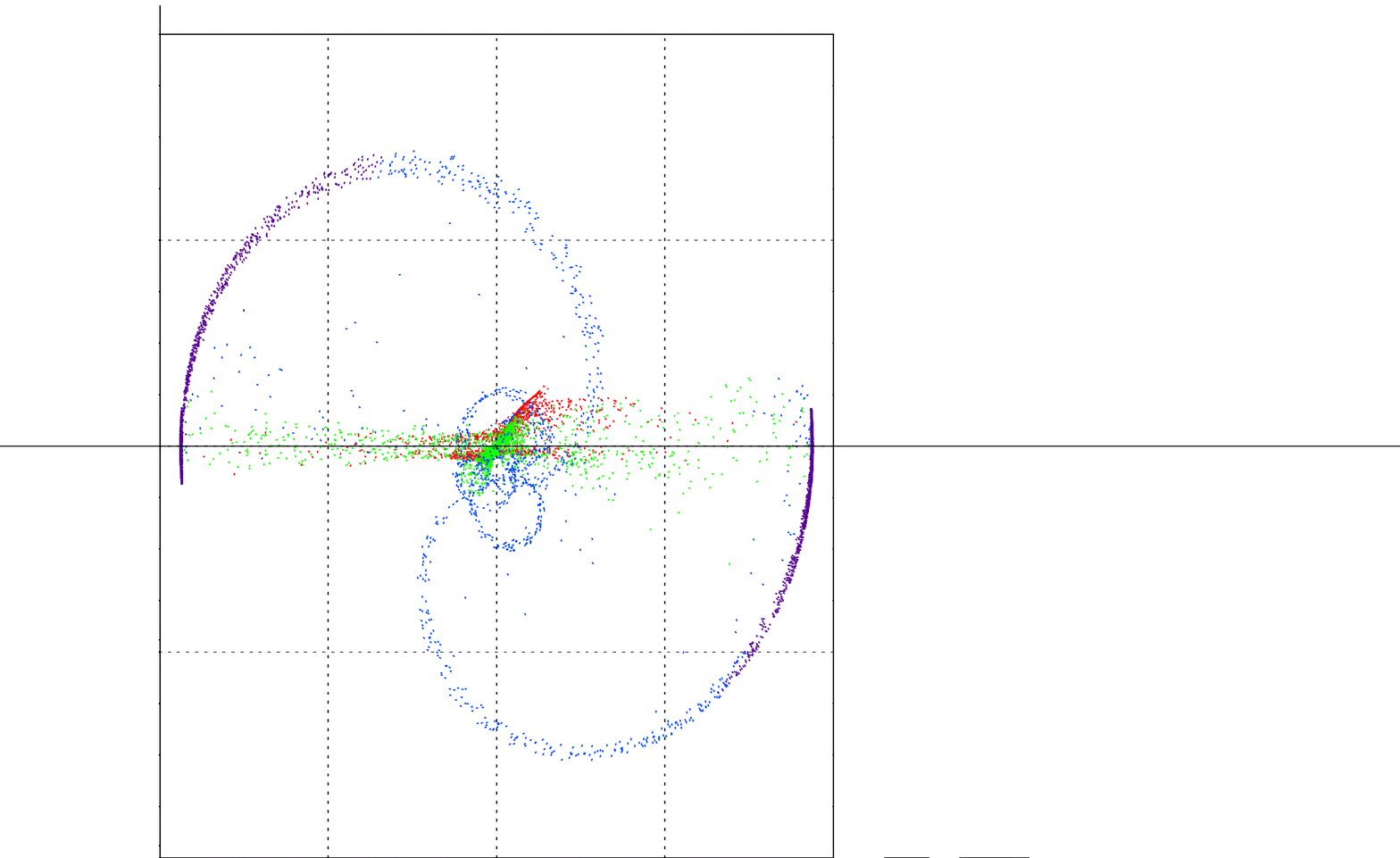}
\caption{(top) Cosmic-ray halo formed 12,000 years after a GRB that took
place at 3 kpc from the center of the Galaxy. For clarity, equal
numbers of cosmic rays are injected per decade, with cosmic rays
color-coded by energy. Cosmic-rays protons with Lorentz factors
$\gamma$ in the ranges $10^7$ -- $10^8$, $10^8$ -- $10^9$, $10^9$ --
$10^{10}$, and $10^{10}$ -- $10^{11}$ are red, green, turquoise, and
purple, respectively. Cosmic-ray neutrons and neutron-decay protons
are dark red, yellow, magenta, and dark blue in the respective energy
ranges. (Bottom) Side view of the cosmic-ray halo when diffusive
scattering is turned off (grid lines are 2 kpc in separation). }
\label{fig2}
\end{figure}
\clearpage

\section{Discussion and Summary}

About once every several hundred million years, the Earth is illuminated
by prompt GRB photon and neutral radiations of sufficient intensity to have
significant effects on the biota through erythema from cascade UV flux
\citep{sw02,ssw04} and destruction of the ozone layer catalyzed by 
the formation of nitric oxide and NO$_y$ ``odd nitrogen" compounds
\citep{rud74,tho95,tho05}. The destruction of the ozone layer causes
the Solar UV fluence to greatly exceed the GRB cascade UV fluence when 
the net effects of the GRB radiation on the atmosphere are considered \citep{tho05}.
Such an event might have been responsible 
for trilobite extinction in the Ordovician era through the destruction
of plankton \citep{mel04}.

If high-energy cosmic-ray production accompanies GRBs, then Galactic
events could also have affected biological evolution due to DNA
radiation damage by the elevated ground-level muon fluxes induced by
$\approx 10^{17}$ eV ($\gamma_8 =\gamma /10^{8}\cong 1$) cosmic-ray
neutrons \citep{dls98}.  The prompt energy fluence of neutrons is
\begin{equation}
\varphi_n(\gamma,r) = m_p c^2\gamma^2 
{dN_n\over d\gamma_n}{\exp(-r/r_n)\over f_b\cdot 4\pi r^2}
\cong  {4\times 10^8 \eta_n{\cal E}_{52}\over f_{-2} 
r_{\rm kpc}^2}\; {\exp(-r_{\rm kpc}/\gamma_8)\over 
\gamma_8^{-1.2}}\;
\;{\rm ergs~ cm}^{-2}\;,
\label{varphi}
\end{equation} 
where an on-axis GRB at a distance of  $r_{\rm kpc}$ kpc 
accelerates $10^{52} \eta_n{\cal E}_{52}$ ergs 
of nonthermal neutrons above $\approx 100$ TeV into a two-sided jet
with beaming factor  $f_{-2}$\%.
The prompt muon number flux due to this type of event is at the level
\begin{equation}
\varphi_\nu(>E_\mu )\cong {4\times 10^{11}\sec\theta\over 
E_\mu^{1.76}({\rm GeV})\gamma_8^{0.44}}
{\eta_n {\cal E}_{52}\over f_{-2} r_{\rm kpc}^2} 
\exp(-r_{\rm kpc}/\gamma_8)\;
\label{varphinu}
\end{equation}
\citep{gai90} for GRBs within $\pi/3$ of the zenith. This value is above the level
$\varphi_\nu(> 3 {\rm ~GeV})\sim 10^{10}$ for 50\% mortality of human
beings. The situation could be worse because of various leptonic and
hadronic radiation pathways in the GRB leading to significant fluxes
of $> 100$ MeV -- TeV radiation and enhanced cascade UV flux \citep{dr02}.

Deflection of charged particles by the Galactic magnetic field means
that cosmic-ray protons arrive after the delay time $\delta t
\simeq 2 r_{\rm L}(\theta - \sin\theta)
\simeq r_{\rm L}\theta^3/3c$ for $\theta \ll 1$, where the characteristic
Larmor radius $r_{\rm L} \cong 0.1 \gamma_8/B_ {\mu{\rm G}}$ kpc is
assumed to be much larger than the source distance, and $\theta$ is a
characteristic deflection angle.  For a GRB 1 kpc away, $\gtrsim
4\times 10^{18}$ eV protons arrive within angle $\theta$ of the source
direction over $\approx 1000 \theta^3 $ years, taking $B_{\mu{\rm
G}}\approx 4$.  Diffusive scattering spreads out the arrival times and
directions compared to this estimate, and implies a ``chirping" behavior
where higher energy particles arrive first. From the preceding expression,
the energy dependence of the time delay for a proton with Lorentz 
factor $10^9\gamma_9$ is given by 
$\Delta t \cong 10^2 d_{kpc}^3B_{\mu{\rm G}}/\gamma_9$ s
for $\gamma_9 \gtrsim B_{\mu {\rm G}} d_{kpc}$, and
$\Delta t \cong 10^2 d_{kpc}^2/\gamma_9$ s for $10^{-2} \lesssim \gamma_9 \lesssim 1$, 
using the form of $\lambda(\gamma)$ from \citet{wda04}.

\begin{figure}[t]
\vskip-0.5in
\includegraphics[scale=0.5]{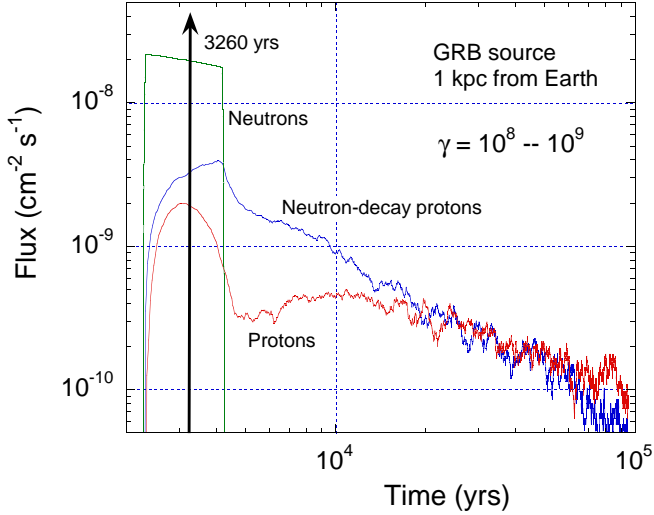}
\vskip-1.75in
\caption{Fluxes of cosmic rays with $10^8 \leq \gamma 
\leq 10^9$ received at Earth from an on-axis 
GRB that occurred 1 kpc away. The fluxes of cosmic-ray neutrons,
neutron-decay protons, and protons are shown by the curves that
decline, rise, and peak during the initial phase.  The prompt flux
actually lasts for some minutes to hours, rather than the 2000 yrs
indicated by the numerical simulation, which reflects the
approximations of the numerical method.  }
\label{fig3}
\end{figure}

Fig.\ 3 shows the time-dependence of the flux of cosmic-ray neutrons and protons
with  Lorentz factors $10^8 \leq \gamma \leq 10^9$
received from a GRB source located one kpc from the Earth. The received 
flux of prompt
neutrons with this range of Lorentz factors would be 
$dN_n/dAdt \cong 1250/[(\theta_j/0.1)^2 t_{dur}({\rm s})]$ cm$^{-2}$ s$^{-1}$,
where $t_{dur}({\rm s})$ is the mean duration of the cosmic-ray emission event.
Because of the numerical method, this duration has been artificially 
increased to  $6.1\times 10^{10}$ s, so that if the actual duration of the event
were 600 s,
the flux would  be $10^8\times$ greater and the duration $10^8\times$
shorter than shown in the figure, giving constant total fluence.
The flux units shown in Fig.\ 3 
apply directly to the cosmic-ray protons and neutron-decay protons, 
and we see that the cosmic-ray fluence in the prompt and extended phases are roughly equal.
The measured energy density of $10^{17}$ -- $10^{18}$ eV cosmic rays is $\approx
10^{-18}$ ergs cm$^{-3}$, so Fig.\ 3 shows that the cosmic ray
energy density at these energies would be $\approx 4$ orders of magnitude
larger than the currently measured energy density during a period of $\approx 10^4$ 
yrs following such an event.

By depositing larger cosmic ray fluence during a much shorter interval, the prompt flux has, 
however, much greater lethality and effect.  The occurrence of the sequence
of extinction events in the Ordovician due to a long lasting $\sim 1$
Myr ice age \citep{mel04} could happen if the prompt blast induced a
long term change in the climate, or if the delayed cosmic rays
induce a glaciation \citep{sha03}.  On-axis events are considerably
more damaging than off-axis events which, though more numerous,
release $\lesssim 10^{17}$ eV cosmic rays that slowly diffuse towards
Earth over periods of thousands of years and longer.

The hypothesis \citep{bie04} that the $\gtrsim 10^{18}$ eV cosmic-ray
excesses detected with the AGASA and SUGAR arrays \citep{hay99} are
cosmic-ray neutrons from a GRB is not, however, supported by our
simulations.  The relativistic blast waves in a GRB accelerate the
highest energy cosmic rays over timescales of weeks or less
\citep{zm04}, and the high-energy neutrons therefore arrive on this
same timescale. Cosmic ray protons with energies $\sim 10^{19}$ eV are
delayed over a timescale $\approx 10000\theta^3/B_{\mu{\rm G}}$ years
from a source at the distance of the Galactic Center. For the SUGAR
excess, which is coincident on the sub-degree ($\theta \lesssim 0.02$)
angular scale with a point source, a GRB would have to take place
within weeks of the observation for cosmic-ray protons to maintain
their direction to the source.  Including the requirement that the GRB
jet was also pointed towards Earth means that an impulsive GRB origin is
excluded because such an event is highly improbable.  The greater
($\approx 10^\circ$) extent of the AGASA excess does not conclusively
exclude a GRB origin, but here the diffuse excess could simply reflect
the greater pathlength for cosmic-ray proton collisions with spiral
arm gas along the Cygnus arm.

Because the SUGAR point source does not admit an impulsive GRB
solution, only cosmic rays from a persistent source, such as a
microquasar, could make such an excess. The hypothesis
\citep{der02,wda04} that GRBs are sources of $\gtrsim 10^{14}$ eV 
cosmic rays is therefore incompatible with such a source.  This
cosmic-ray origin hypothesis will soon be tested by results from the
Auger Observatory\footnote{www.auger.org} to confirm this source. If
the source is real, then the GRB/cosmic-ray model is
incomplete. Searches for neutron $\beta$-decay radiation in recent
galactic GRBs \citep{ikm04} and around galaxies that host GRBs
\citep{der02} provide further tests of the hypothesis that high-energy
cosmic rays are accelerated by GRBs.

The Earth resides on the inner edge of a spiral arm, and not in an OB
association where high-mass stars, and therefore GRBs, are usually
found. The greater likelihood for intense irradiation events by GRBs
excludes OB associations from the galactic habitable zone \citep{lfg04}
except for planets with very thick ($\gtrsim 1000$ gm cm$^{-2}$)
atmospheres \citep{ssw04} needed to limit the radiation effects.

\vskip0.1in
{\bf Acknowledgements:} This work is supported by the Office of Naval
Research and the NASA {\it Gamma Ray Large Area Space Telescope}
(GLAST) program. We thank A.\ Atoyan, J.\ D.\ Kurfess, A.\ L.\ Melott, 
K.\ E.\ Mitman, and S.\ D.\ Wick for discussions and comments on the 
manuscript. We also thank the referees for their reports, including 
the request to clarify the applicability of the test-particle limit.

\end{document}